\documentclass{aa}
\usepackage{graphicx}
\usepackage{natbib}
\usepackage{epsfig}
\def\ut#1{\mathop{\vtop{\ialign{##\crcr
     $\hfil\displaystyle{#1}\hfil$\crcr\noalign
     {\kern1pt\nointerlineskip}\hbox{$\hfil\sim\hfil$}\crcr
     \noalign{\kern1pt}}}}}

\def\undersymbol#1#2{\mathop{\vtop{\ialign{##\crcr
     $\hfil\displaystyle{#2}\hfil$\crcr\noalign
     {\kern1pt\nointerlineskip}\hbox{$\hfil#1\hfil$}\crcr
     \noalign{\kern1pt}}}}}
\def\arcsec{^{\prime\prime}}
\def\arcmin{^{\prime}}
\def\degr{^0}

\usepackage{graphicx}

\begin{document}

\title{Planck view of the M82 galaxy}
       \author{V.G. Gurzadyan\inst{1,2}, F. De Paolis\inst{3,4}, A.A. Nucita\inst{3,4}, G. Ingrosso\inst{3,4},  A.L. Kashin\inst{2}, H.G. Khachatryan\inst{2}, S. Sargsyan\inst{2}, G. Yegorian\inst{2}, Ph. Jetzer\inst{5}, A. Qadir\inst{6}, \and D. Vetrugno\inst{7}}
              \institute{SIA, Sapienza University of Rome, Rome, Italy 
              \and Center for Cosmology and Astrophysics, Alikhanian National Laboratory and and Yerevan State University, Yerevan, Armenia 
              \and Dipartimento di Matematica e  Fisica ``E. De Giorgi'', Universit\`a del Salento, Via per Arnesano, I-73100, Lecce, Italy  
              \and INFN, Sez. di Lecce, Via per Arnesano, I-73100, Lecce, Italy 
\and
Physik-Institut, Universit\"at
Z\"urich, Winterthurerstrasse 190, 8057 Z\"urich, Switzerland
\and
School of Natural Sciences,
National University of Sciences and Technology, Islamabad,
Pakistan
\and 
Department of Physics, University of Trento, I-38123 Povo, Trento, Italy and 
TIFPA/INFN, I-38123 Povo,  Italy
}

   \offprints{F. De Paolis, \email{francesco.depaolis@le.infn.it}}
   \date{Submitted: XXX; Accepted: XXX}

 \abstract{Planck data towards the galaxy M82 are analyzed in the 70, 100 and 143 GHz bands. A substantial north-south and East-West  temperature asymmetry is found, extending up to $1\degr$ from the galactic center. Being almost frequency-independent, these temperature asymmetries are indicative of a Doppler-induced effect regarding the line-of-sight dynamics on the halo scale,
the ejections from the galactic center and, possibly, even the tidal interaction with M81 galaxy. The temperature asymmetry thus acts
as a model-independent tool to reveal the bulk dynamics in nearby edge-on spiral galaxies, like the Sunyaev-Zeldovich effect for
clusters of galaxies. 
 }


   \authorrunning{De Paolis et al.}
   \titlerunning{Planck view of the M82 galaxy}
   \maketitle
%

\section{Introduction}
M82 (NGC 3034, also known as the Cigar galaxy) is the largest galaxy of the M81 group in the Ursa Major constellation.  It is nearly edge-on (its inclination angle to the line of sight is $i\simeq 80\degr$), and it is the closest galaxy that hosts a starburst nucleus, being at a distance of only 3.5 Mpc \citep{dalcanton2009}. \footnote{At the M82 distance, $1\arcmin$ corresponds to about 1 kpc.} Originally classified as a dwarf irregular type-II galaxy \citep{sandagebrucato1979}, M82 has been more recently reclassified as a late type SBc galaxy after the discovery of a bar (oriented almost along the major axis of the galaxy and with about $1\arcmin$ length) and arm structure in its disk 
\citep{telesco1991,achtermann1995,mayya2005}. The coordinates of the M82 center are R.A. (J2000)= $09^{\rm h} 55^{\rm m} 52.7^{\rm s}$, Dec. (J2000)=$69\degr 40\arcmin 46\arcsec$ (corresponding to galactic coordinates $l=141.409498\degr$ and $b=40.567579\degr$). It is one of the most studied galaxies after the Milky Way  (see, e.g., \citealt{markarian1962}) and has been extensively observed at all wavelengths, from  the radio band to high energies (see, e.g., \citealt{hutton2014}). In the IR, M82 is the brightest galaxy in the sky, and shows an outwardly expanding gas \citep{hutton2014}, which is probably being driven out by the combined emerging particle winds of many stars, together creating a galactic superwind (see also \citealt{gandhi2011}). Stellar dynamics measurements of the innermost bulge region of the central M82 galaxy show that it harbors a supermassive black hole with a mass $\simeq 3\times 10^7$ M$_{\odot}$ \citep{gaffney1993}.

The M82 rotation curve has been derived in several studies using both stellar and gaseous tracers up to a radius of  about $170\arcsec$  ($\simeq 2.9$ kpc). \cite{mayyal1960} and \cite{goetz1990} obtained a rotation curve using optical emission and absorption lines with a peak value of about 130 km s$^{-1}$, whereas \cite{sofue1998} used the CO and H I lines to find  a peak value of about 190 km s$^{-1}$. Both analyses find a steeply rising curve, peaking at about $25\arcsec$ from the galaxy center and rapidly declining in a Keplerian fashion beyond $25\arcsec$, likely indicating that  M82's extended disk mass is missing, which is very peculiar and exceptional for a disk galaxy. More recently, \cite{greco2012} have presented a K-band spectroscopic study and measured the M82 rotation curve out to about $4\arcmin$ ($\simeq 4$ kpc) on both the eastern and western sides of the galaxy. At variance with the previous studies, it has been found that the stellar rotation velocity is flat on scales of $1-4$ kpc  (with only a slight decline on the eastern side) with a value of $\simeq 110-120$ km s$^{-1}$ out to 4 kpc, thus implying an M82 total dynamical mass of $\sim 10^{10}$ M$_{\odot}$. 

As  shown by  \cite{depaolis2011,depaolis2014,depaolis2015}, CMB data offer a unique opportunity to  study the  large-scale temperature asymmetries far beyond the size typically accessible with other tools toward nearby  astronomical systems.  We study the M82 galaxy by using {\it Planck} data within a galactocentric radius of $1\degr$.

\section{Planck data analysis and results}
The sensitivity, angular resolution, and frequency coverage of its detectors make {\it Planck}  a powerful instrument for both cosmology and galactic and extragalactic astrophysics \citep{planck2015a}. Here we use the data of {\it Planck} 2015 release \citep{planck2015a} in the bands at 70 GHz of the Low Frequency Instrument (LFI), and in the bands at 100 and 143 GHz of  the High Frequency Instrument (HFI). For a review of {\it Planck} results and instrument characteristics, we refer the interested reader to  \cite{burigana2013}, among others.
{\it Planck}'s resolution is $13\arcmin$, $9.6\arcmin$, and $7.1\arcmin$ in terms of FHWM at 70, 100, and 143 GHz bands, respectively,
 and frequency maps \citep{planck2015b} are provided in CMB temperature at resolution corresponding to $N_{side}$=2048 in HEALPix scheme \citep{gorski2005}.

\begin{figure}[h!]
 \centering
  \includegraphics[width=0.50\textwidth]{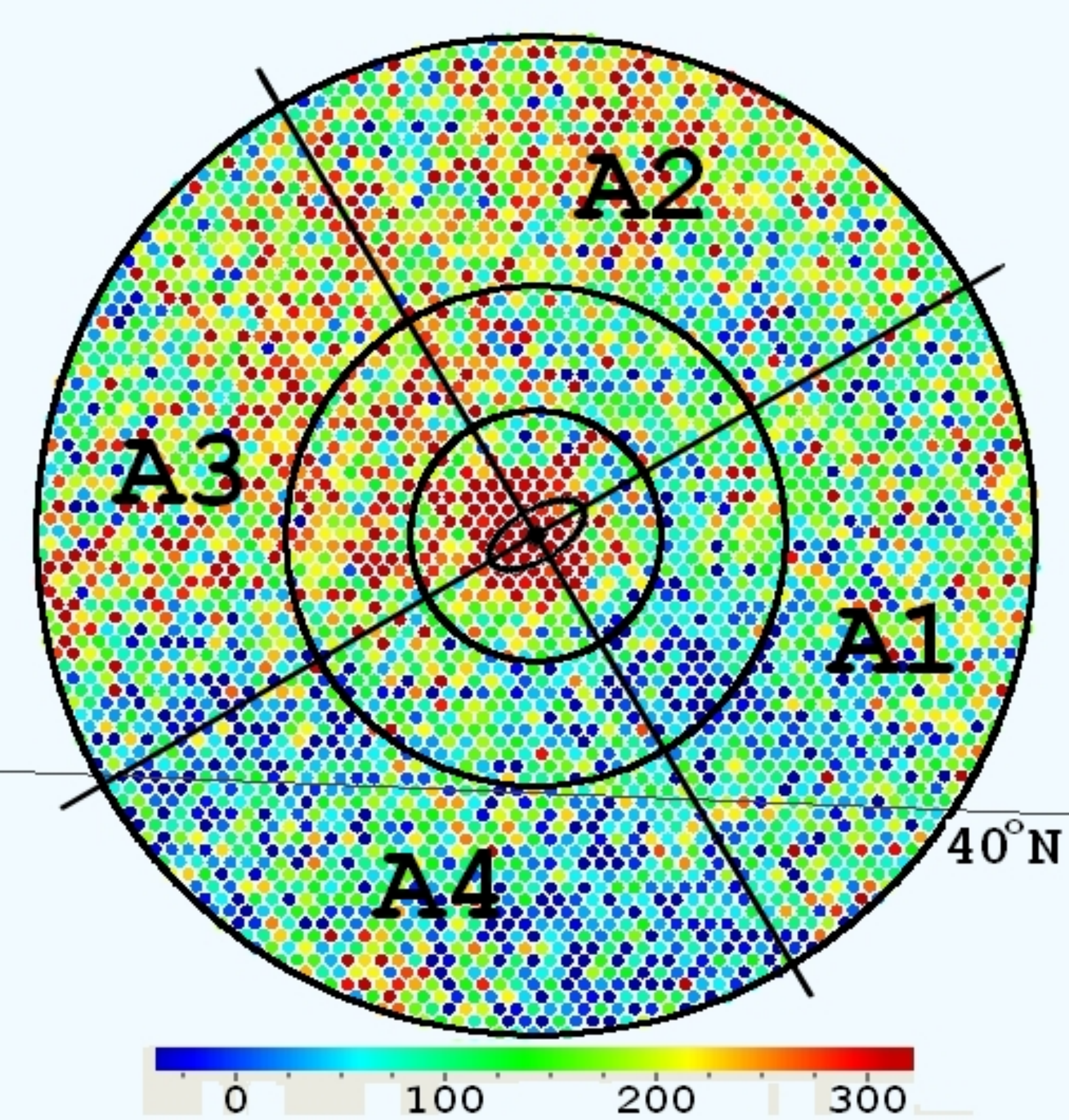}
 \caption{The {\it Planck} field toward the M82  galaxy in the 70GHz band. The pixel color gives the temperature excess with respect to the mean CMB temperature in $\mu$K. The galactocentric radii of the circles are $15\arcmin$, $30\arcmin$ and $1\degr$, respectively. The M82 galaxy is indicated by the inner ellipse with major  and minor axes of  $10.73\arcmin$ and $5.02\arcmin$, respectively. The analysis presented in Section 2 is performed within the quadrants indicated as A1, A2, A3, and A4. The line corresponding to Galactic latitude $40\degr$ North is also shown.} \label{fig1}
 \end{figure}

To study the CMB data toward the M82 galaxy,  the considered region of the sky (shown in Fig. \ref{fig1} in the {\it Planck} band at 70 GHz) has been divided into  four quadrants indicated as A1, A2, A3, and  A4. The inner ellipse indicates the M82 galaxy extension in the optical band, while the circles are at galactocentric radii $15\arcmin$, $30\arcmin$, and $1\degr$, respectively.

\begin{figure}[h!]
 \centering
  \includegraphics[width=0.46\textwidth]{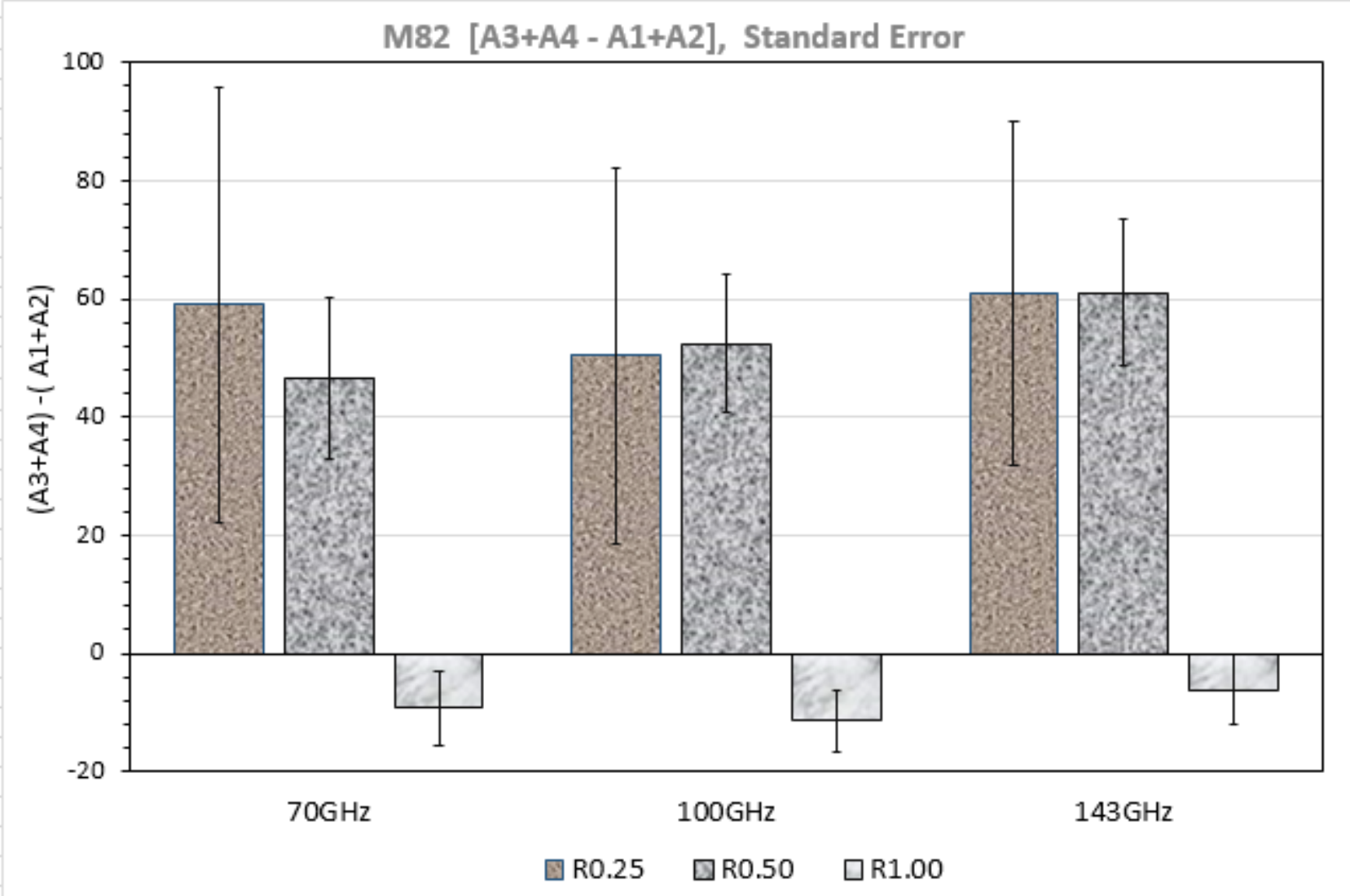}
    \includegraphics[width=0.44\textwidth]{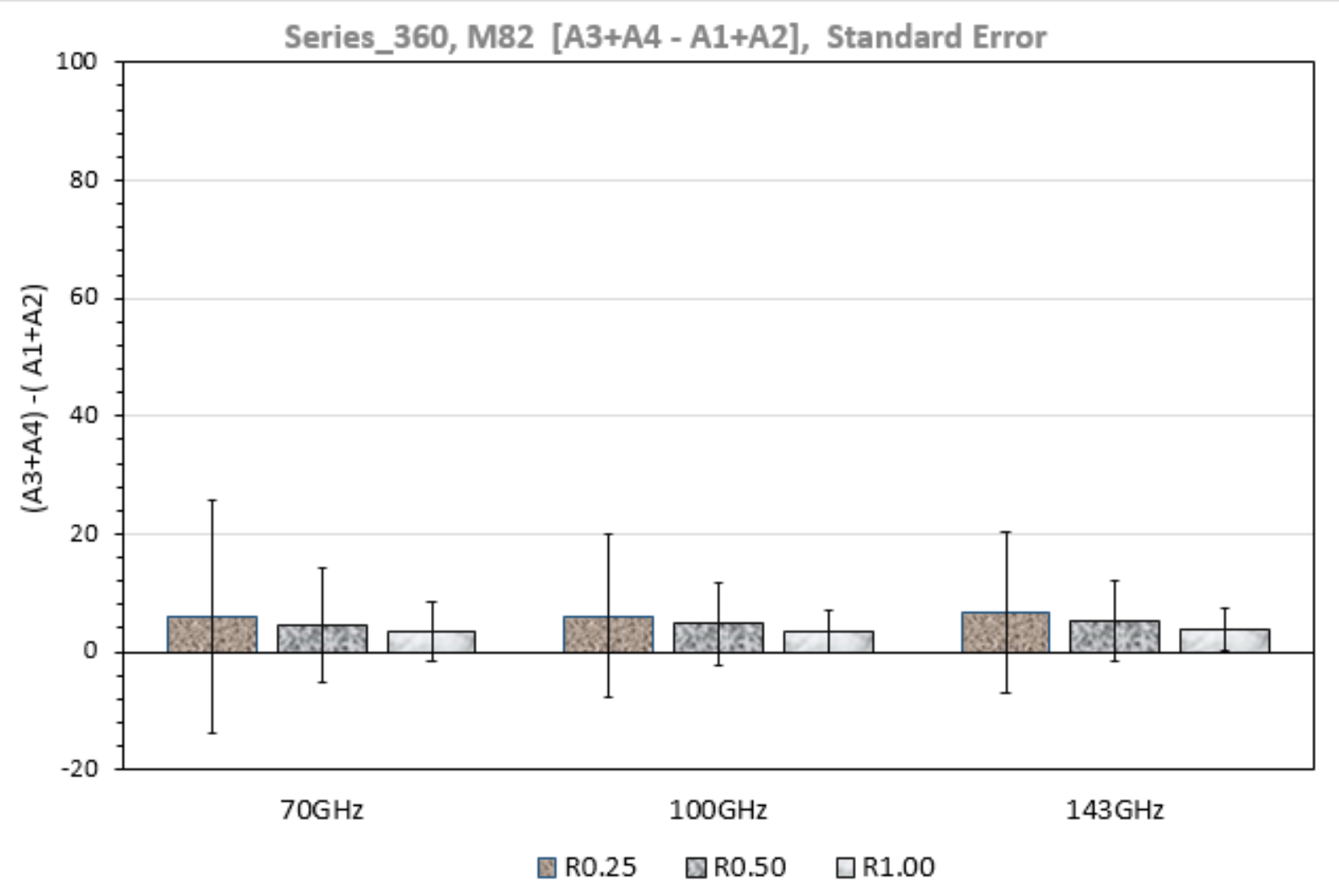}
 \caption{Upper panel: the excess temperature toward M82 in $\mu$K (with the standard errors) of the A3+A4 region with respect to the A1+A2 region in the three considered {\it Planck} bands up to galactocentric radii of  $15\arcmin$ (R0.25), $30\arcmin$ (R0.50), and $1\degr$ (R1.00). Bottom panel: the same for the 360 control fields with the same geometry (shown in Fig. \ref{fig1}) equally spaced at one degree distance to each other in Galactic longitude and at the same latitude as M82.}
 \label{fig2}
 \end{figure}

The mean
temperature excess (with respect to the mean CMB temperature) in $\mu$K in each of the indicated 
regions was obtained in each {\it Planck} band at 70, 100, and 143 GHz with
the corresponding standard error \footnote{The standard error has been 
calculated as the standard deviation of the excess temperature
distribution divided by the square root of the pixel number in
each region. We have verified that within the errors, the sigma
values calculated in that way are consistent with those evaluated
by using the covariance matrix obtained by a best-fitting
procedure with a Gaussian to the same distribution.}.
As one can see from the upper panel of Fig. \ref{fig2}, the A3+A4 region is systematically hotter than the A1+A2 region with an excess temperature almost constant within $15\arcmin$ and $30\arcmin$ in the three {\it Planck} bands at 70, 100, and 143 GHz. The detected excess with respect to the minor axis of the M82 galaxy is therefore aligned along the rotational direction,  indicating a Doppler induced effect modulated by the spin of the galaxy. However, the temperature asymmetry completely disappears if one considers the data within $1\degr$, and this indicates that the temperature asymmetry between the two M82 sides is substantial only up to about $30-40$ kpc.

  \begin{figure}[h!]
 \centering
  \includegraphics[width=0.46\textwidth]{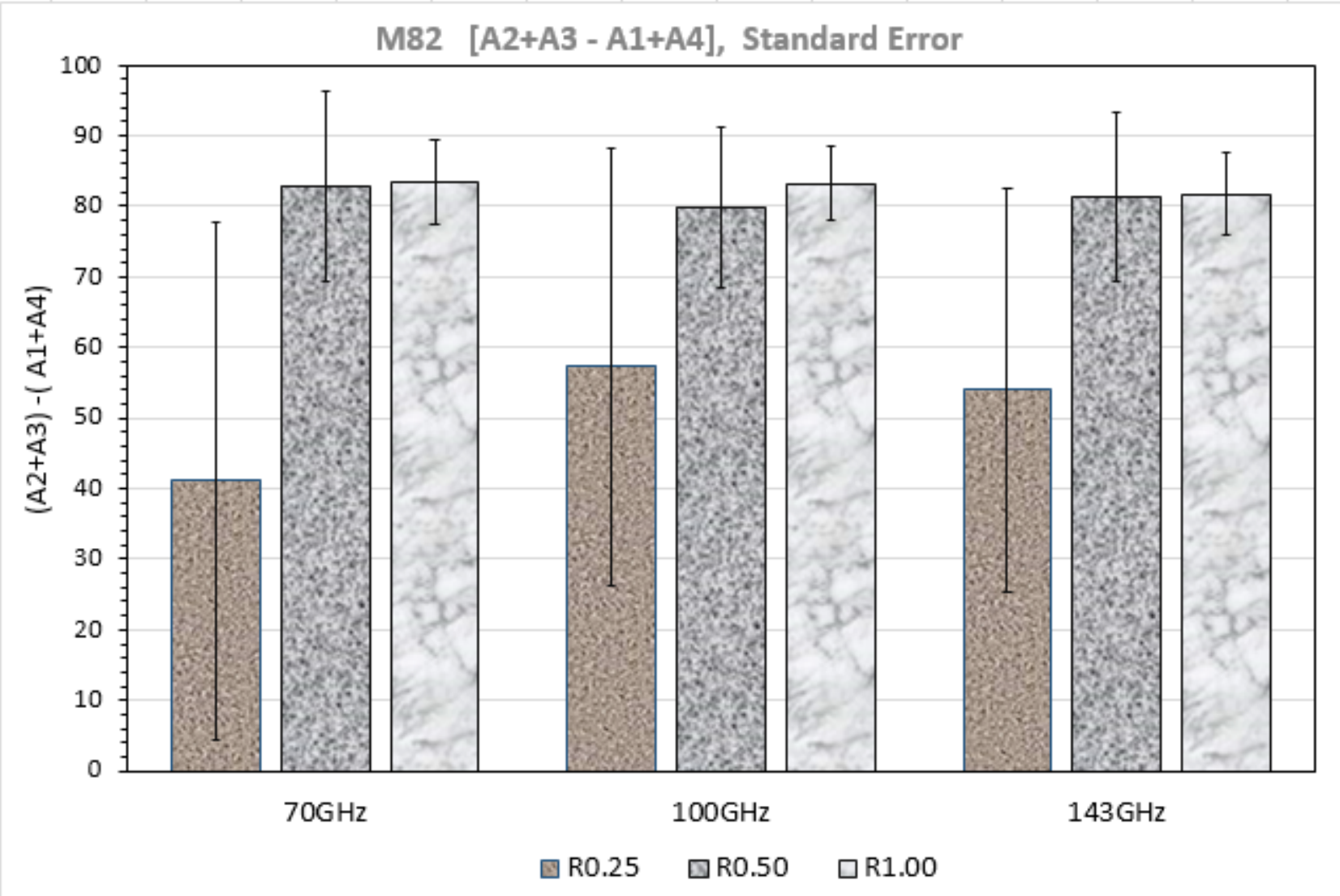}
    \includegraphics[width=0.44\textwidth]{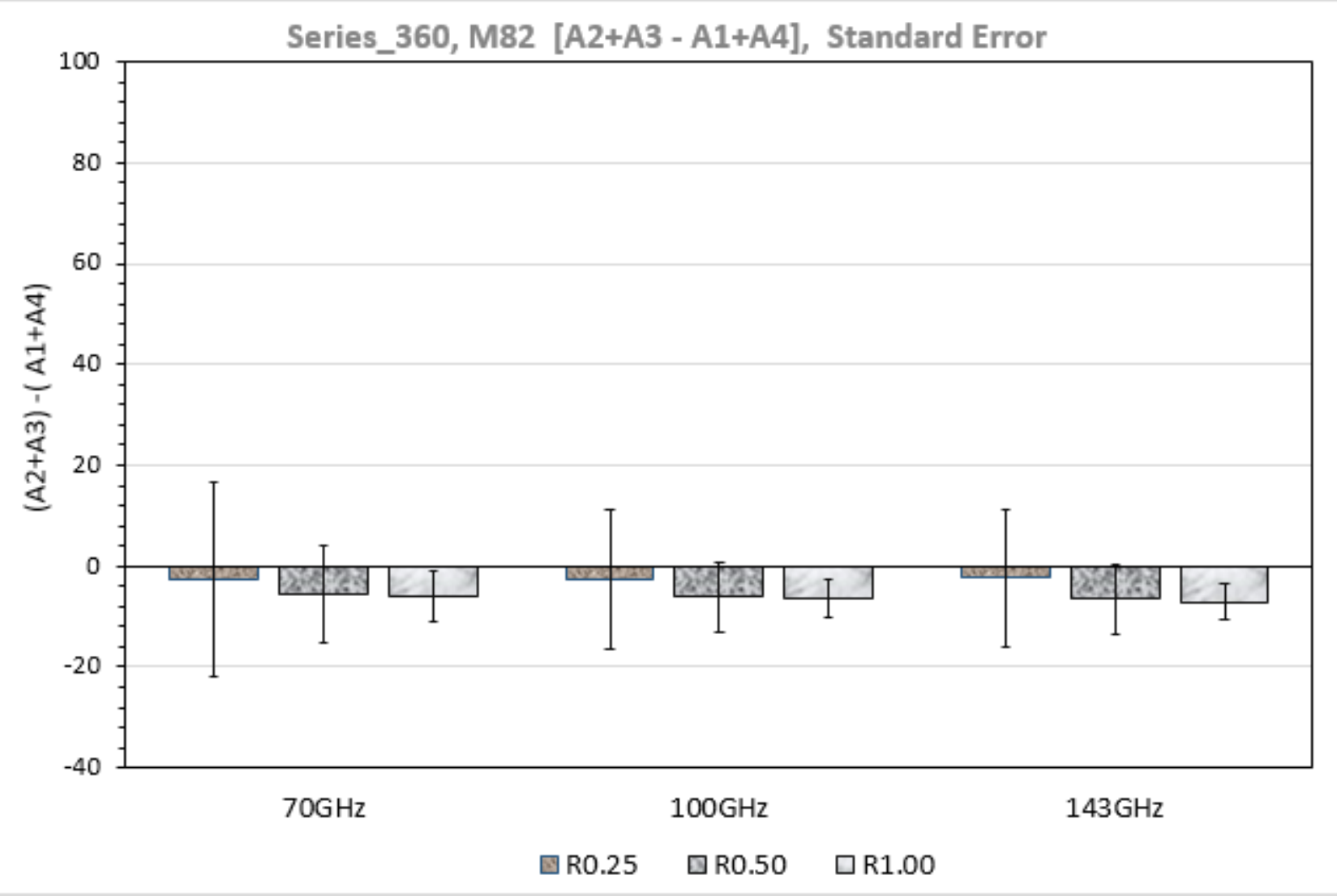}
 \caption{Upper panel: the excess temperature in $\mu$K (with the standard errors) of the A2+A3 region with respect to the A1+A4 region  in the three considered {\it Planck} bands up to galactocentric radii of  $15\arcmin$, $30\arcmin$, and $1\degr$ toward the M82 galaxy. Bottom panel: the same for the 360 control fields  as in Fig. \ref{fig2}.}
 \label{fig3}
 \end{figure}

To test whether the temperature asymmetry we see is real or can be explained by a random fluctuation of the CMB signal (which  is very patchy, especially on angular scales of $0.5\degr -1\degr$), we  considered 360 control field regions with the same shape as shown in Fig. \ref{fig1} at the same Galactic latitude of M82, but at  $1\degr$ longitude from each other. For each region, we determined the excess temperature profile and calculated the average profile and corresponding standard deviation. The results  are  shown in the histogram in the bottom panel of  Fig. \ref{fig2}. As one can see, the temperature asymmetry in the 360 control fields in the three considered {\it Planck} bands is null (within the standard errors), when integrating within  $15\arcmin$,  $30\arcmin$, and $1\degr$. Here we note that the standard errors within $15\arcmin$ toward the M82 galaxy are quite large, as expected, owing to the small number of pixels. Therefore, in the three considered bands and within R0.25  ($15\arcmin$), the temperature asymmetry excess is detected within a rather low confidence level (slightly more than 1 standard deviation). The situation becomes better when considering the intermediate circle at $30\arcmin$ (R0.50).  For example, in this case in the 100 GHz band, we detect a temperature excess of $50\pm 10$ $\mu$K, while the control fields give a temperature excess $\leq 12$ $\mu$K. Therefore, in this region the temperature asymmetry is detected with a confidence level of about $3\sigma$. 

Next, we consider the temperature asymmetry with respect to the major axis of the M82 galaxy. In the upper panel of Fig. \ref{fig3} we give the CMB temperature asymmetry of the A2+A3 regions with respect to the other two regions. As in Fig. \ref{fig2} we also give the temperature asymmetry in the 360 control fields, which is consistent with zero in the three considered bands  and within $15\arcmin$ (R0.25), $30\arcmin$ (R0.50), and $1\degr$ (R1.00). As one can see, the detected temperature asymmetry is substantial (since $\simeq 40$ $\mu$K to $\simeq 80$ $\mu$K) in all the {\it Planck} bands within at least $1\degr$ ($\simeq 60$ kpc) of  the M82 center. The very fact that the detected temperature asymmetry is almost frequency independent is a clear and strong indication of an effect of the galaxy rotation. 

Before closing this section, we stress here that the procedure we followed to test the reliability of the detected temperature asymmetry, that is, of considering the 360 control regions, is more reliable than simulating the CMB sky maps in each band being considered. While the latter methodology of generating sky maps to estimate the error bars is mandatory  when dealing with the whole sky (as in cosmological studies), in our case we are considering only fairly small regions of the {\it Planck} sky maps, and the adopted procedure is more reliable since it avoids the  ambiguities possibly involved in the simulations.

\section{Conclusions and further perspectives}

We have considered {\it Planck} 2015 release data in the bands at 70, 100, and 143 GHz  and detected both a east-west and a north-south  temperature asymmetry toward the M82 galaxy. We detect a temperature asymmetry of the A3+A4 versus the A1+A2 regions (that is aligned with respect to the M82 rotation axis). This is  present in all the considered {\it Planck} bands  up to a galactocentric distance of  $30$ kpc.
Even more robust is the temperature asymmetry of the A2+A3 versus the A1+A4 regions, which extends up to at least 60 kpc, since it is oriented as the M82 outflow. 

The very fact that the detected temperature asymmetry is almost frequency independent and is approximately in the galaxy spin direction (although probably tilted toward the M82 outflow direction) is a rather clear indication of a Doppler-induced effect modulated by the spin of the M82 galaxy. Since the temperature asymmetry is detected far beyond the optical extension of the galaxy, it has to bring the imprint of the M82 halo, whose rotation axis is probably tilted in the west direction with respect to the spin axis of the disk. This conclusion is reinforced by the observation that the rotation axis of the ionized emission-line gas is offset from the stellar rotation axis and the photometric major axis by an angle of about 
$12\degr$ \citep{westmoquette2009}. This has been interpreted  as possibly resulting from the interaction of M82 with the M81 galaxy (which is placed approximately in the A3 region in Fig. \ref{fig1} at about $40\arcmin$ distance from the M82 center) and/or with NGC 3077 (which lies in the A2 region but beyond the R1.00 circle)  in the past. 

We also mention that the rotation of the M82 galaxy up to galactocentric distances of a few kpc has also been studied recently  through the kinematics of planetary nebulae \citep{johnson2009}. The individual radial velocities of these planetary nebulae not only agree with the rotation curve measured with other methods (see the discussion in the Introduction), but moreover, the nebulae at high Galactic latitude  (away from the M82 disk) show a clear rotation signature, in agreement with our results.

Before  closing,  we would like to note that the {\it Planck} Early Release Compton Source Catalogue has been used, together with WMAP (Wilkinson Microwave Anisotropy Probe) data  by \cite{peel2011}, to derive the continuum spectrum of the M82 galaxy. Then, by a least-squares fit to the obtained spectrum by a combination of synchrotron, free-free and thermal dust models, the authors have found evidence of AME (anomalous microwave emission) from this galaxy. However, in a different way from the analysis presented here, the study by \cite{peel2011} is limited to the optical extension of M82 (essentially the 1-2 {\it Planck} pixels of the ellipse at the center of Fig. \ref{fig1}),  while our study is a  ``spatial'' analysis of the region around M82, and, hopefully, brings information from its halo, whose presence has hitherto been expected.
Actually, regardless of the emission mechanism  that may contribute to the signal at CMB wavelengths, \footnote{These mechanisms are ($i$) free-free emission, ($ii$) synchrotron emission, ($iii$) anomalous microwave emission (AME) from dust grains, ($iv$) the kinetic Sunyaev-Zel'dovich (kSZ) effect, and ($v$) cold gas clouds populating the outer galaxy regions (as first proposed, in the context of the M31 halo, by \citealt{depaolis1995}).}
the Doppler-induced temperature asymmetry is given by
\begin{equation}
\frac{\Delta T}{T}=2\tau_{\rm eff}\frac{v}{c}\sin i
\label{eq1}
\end{equation}
where $v$ is the galaxy rotation velocity, $i$  the inclination angle of the galaxy rotation axis with respect to the line of sight \footnote{In the case of the galactic disks, it can be roughly estimated as $i=\arccos(b/a)$, where $b$ and $a$ are the minor and major axes of the galaxy.}, and $\tau_{\rm eff}$  the effective optical depth proper of the particular emission mechanism responsible for the temperature asymmetry. For example,  for the kSZ case  $\tau_{\rm eff}$ is the projected optical depth due to Thomson scattering of the CMB photons on free electrons (see, e.g., \citealt{mak2011}), while $\tau_{\rm eff}=\bar{\tau} S$ in the case of the cold clouds mechanism (where $\bar{\tau}$ is the frequency averaged cloud optical depth and $S$  is the cloud filling factor, as in  \citealt{depaolis1995}). 
As a matter of fact and keeping in mind the object of the present study, mechanisms ($i$), ($ii$), and ($iv$), as detailed in  footnote 3, require the presence of hot plasma in the M82 halo -- which is indeed observed up to about $6$ kpc \citep{bregman1995} along the minor axis but might also be more extended as shown by Suzaku observations \citep{tsuru2007} -- while mechanisms  ($iii$) and ($v$)  only involve an extended population of cold gas and/or dust, which is observed as well 
 \footnote{We note that $H_2$ knots and filaments in the M82 halo have been detected  even beyond $5$ kpc \citep{veilleux2009,beirao2015}.}.

The importance of the methodology we propose, which is applied to the M82 case, is that in spite of its simplicity, it may allow us to consistently estimate the dynamical mass $M_{\rm dyn}$ (contained within a certain galactocentric distance $R$) of the considered galaxy once the temperature asymmetry has been quantified.  In fact,
\begin{equation}
M_{\rm dyn}(<R)=\frac{v^2}{G}R=700\left(\frac{\Delta T_{\mu K}}{\tau_{\rm eff}\sin i }\right)^2R_{\rm kpc}~M_{\odot}
\label{eq2}
\end{equation}
which directly envisages a lower limit to  $M_{\rm dyn}$ (obtained for  $\tau_{\rm eff}=1$) that, in the case of the M82 galaxy, turns out to be ($\Delta T\simeq 80$ $\mu$K and $R\simeq 60$ kpc) $M_{\rm dyn}\geq 2.8\times 10^8~M_{\odot}$. This is consistent with the observations and can be further improved by considering the details regarding the emission mechanisms involved (that is estimating $\tau_{\rm eff}$). The contribution to the revealed temperature asymmetries can be on  different scales from the matter outflow from the galactic center, the halo rotation, and the stripped matter due to M81's interaction.
In general, our method applied to nearby edge-on spirals reveals the galactic halo bulk  dynamics on a rather large scale in a model-independent way. In this sense it resembles the SZ effect, which on galactic scales cannot work since the $e^-$  temperature is not high enough to produce a substantial effect. The importance of the halo's traced parameters (scale, rotation) by this method is also obvious  for dark matter and cosmological reasons, especially if complemented by other dynamical and bulk motion information \citep{rg}. Used in synergy with other data, this may  provide unique keys to studying the bulk dynamics, the outflow ejection processes, and thus the evolution of galactic systems.

\begin{acknowledgements}
{We acknowledge the use of {\it Planck} data in the Legacy Archive for
Microwave Background Data Analysis (LAMBDA) and HEALPix
\citep{gorski2005} package. FDP, AAN, and GI acknowledge the support by the INFN
project TAsP, and PJ acknowledges support from the
Swiss National Science Foundation. 
One of us (AQ) is grateful for hospitality to the DST Centre of Excellence in Mathematical \& Statistical Sciences of the University of the Witwatersrand, Johannesburg, South Africa.}
\end{acknowledgements}


\end{document}